\documentclass[aps, reprint, nofootinbib, longbibliography, superscriptaddress, floatfix, preprintnumbers]{revtex4-2}

\usepackage{graphicx}
\usepackage{dcolumn}
\usepackage{footmisc}

\usepackage{bm}
\usepackage{hyperref}
\usepackage{xcolor}
\usepackage{float,amsmath}

\usepackage{xspace}
\newcommand{\rg}{\,R$_g$\xspace}

\newcommand{\zdiss}{\,$z_{\rm diss}$\xspace}
\newcommand{\fsc}{\,$f_{\rm sc}$\xspace}
\newcommand{\ngc}{NGC~1068\xspace}
\newcommand{\gr}{$\gamma$-ray\xspace}
\newcommand{\grs}{$\gamma$\,rays\xspace}

\newcommand{\pg}{p$\gamma$\xspace}
\newcommand{\blj}{\texttt{BHJet}\xspace}
\newcommand{\hadjet}{\texttt{HadJet}\xspace}

\begin{document}
\setlength{\abovedisplayskip}{5pt}
\setlength{\belowdisplayskip}{5pt}
\setlength{\abovedisplayshortskip}{5pt}
\setlength{\belowdisplayshortskip}{5pt}

\preprint{}

\title{Axion-like particle limits from multi-messenger sources}

\author{Ariane Dekker}
\email{ahdekker@uchicago.edu}
\affiliation{Kavli Institute for Cosmological Physics, The University of Chicago, Chicago, IL 60637 USA}
\author{Gonzalo Herrera}
\email{gonzaloh@mit.edu}
\affiliation{Center for Neutrino Physics, Department of Physics, Virginia Tech, Blacksburg, VA 24061, USA}
\affiliation{Department of Physics and Kavli Institute for Astrophysics and Space Research,
Massachusetts Institute of Technology, Cambridge, MA 02139, USA}
\affiliation{Department of Physics \& Laboratory for Particle Physics and Cosmology,
Harvard University, Cambridge, MA 02138, USA}
\author{Dimitrios Kantzas}
\email{kantzas@lapth.cnrs.fr}
\affiliation{LAPTh, CNRS,  USMB, F-74940 Annecy, France}
\affiliation{New York University Abu Dhabi, PO Box 129188, Abu Dhabi, UAE \\
Center for Astrophysics and Space Science (CASS), New York University Abu Dhabi, PO Box 129188, Abu Dhabi, UAE}

\begin{abstract}
High-energy neutrino observation from the Seyfert galaxy \ngc offers new insights into the non-thermal processes of active galactic nuclei. 
Simultaneous \grs emitted by such sources can possibly oscillate into axion-like particles (ALPs) when propagating through astrophysical magnetic fields, potentially modifying the observed spectrum. 
To probe for ALP-induced signals, a robust understanding of the emission processes at the source is necessary. 
In this work, we perform a dedicated multi-messenger analysis by modeling a jet in the innermost vicinity of the central supermassive black hole of \ngc. We model in particular the neutrino and \gr emission originating in lepto-hadronic collisions between jet accelerated particles and background particles from the corona, reproducing both the Fermi-LAT and IceCube data. These source models serve as a baseline for ALP searches, and we derive limits on the ALP-photon coupling by marginalizing over motivated ranges of astrophysical parameters. 
We find $g_{a\gamma} \lesssim 7 \times 10^{-11}$GeV$^{-1}$ for $m_a \lesssim 10^{-9}$ eV. 
These limits may be weaker than existing constraints, but they demonstrate the potential of multi-messenger observations to probe new physics. 
We conclude by discussing how additional upcoming multi-messenger sources and improved observational precision can enhance ALP sensitivity. 
\end{abstract}

\maketitle

\section{Introduction}
Axion-like particles (ALPs) are light pseudoscalar bosons which arise in several extensions of the Standard Model, and can comprise the observed relic abundance of dark matter in the Universe ~\cite{Peccei:1977hh,Weinberg:1977ma,Preskill:1982cy,Abbott:1982af,Dine:1982ah,Marsh_2016,Arvanitaki_2010}. Unlike the QCD axion, ALPs cannot resolve the strong CP problem, thereby spanning a wider parameter space and offering a richer phenomenology than the conventional axion. A generic feature of ALPs and QCD axions is their coupling to a two-photon vertex, which in the presence of an external magnetic field can lead to ALP-photon oscillations \cite{Raffelt_stodoslky}. 

The ALP-photon coupling can modulate observations of \gr spectra from astrophysical sources with strong magnetic fields. Previous works have used this feature to set stringent limits on ALPs using multi-wavelength observations from, for instance, Active Galactic Nuclei (AGN) with external magnetic fields from their jets or the intracluster medium surrounding the AGN~\cite{Hooper:2007bq,Reynolds_2020,Sisk_Reyn_s_2021,jacobsen2022constrainingaxionlikeparticleshawc,HESS:2013udx,Tavecchio:2012um}, neutron stars~\cite{Dev:2023hax,lecce2025probingaxionlikeparticlesmultimessenger,Diamond:2023cto}, supernovae~\cite{Calore_2022,candón2025freshlookdiffusealp,Meyer:2016wrm}, galaxy clusters~\cite{chan2021constrainingaxionphotoncouplingusing,Schlederer_2016,Abe_2024} and using the magnetic field of the Milky Way's galactic halo~\cite{Simet:2007sa,Conlon_2014,Carenza_2021,Simet:2007sa,Eckner:2022rwf,Mastrototaro:2022kpt}. 

Probing ALPs using modulated \gr spectra from astrophysical sources requires a robust understanding of the intrinsic astrophysical emission processes. Multi-wavelength and multi-messenger observations help to reduce degeneracies between potential ALP signals and conventional astrophysical emission. 
For instance, astrophysical \gr attenuation typically leads to a lower energy flux enhancement, while ALP-induced attenuation is expected to suppress the flux across all wavelengths. 
Observing the spectral flux across all energy bands can therefore help identify the origin of the attenuation. Moreover, if the \gr emission is produced from hadronic interactions, it is accompanied by a high-energy neutrino flux. Combining neutrino and \gr data therefore gives insights into dominant emission processes and might point towards additional mechanism such as ALP-induced effects. 

IceCube has observed several high-energy (TeV-PeV) neutrinos in the past decade that might be associated with AGN. For through-going muon neutrinos, IceCube achieves a median angular resolution of $\sim 1^{o}$ at 1 TeV~\cite{IceCube:2018ndw}. The first neutrino event that was found to be correlated with electromagnetic emission came from the flaring blazar TXS~0506+056 at $3.5\sigma$ significance~\cite{IceCube:2018dnn,IceCube:2018cha}. More recently, the IceCube collaboration reported an excess of $79^{+22}_{-20}$ neutrino events associated to nearby Seyfert type-2 galaxy \ngc at $4.2\sigma$ level of significance~\cite{IceCube:2022der}. Seyfert galaxies are among the most abundant type of AGN at low redshifts~\cite{Ho:2008rf}. 
Interestingly, no emission has been observed in the TeV-range by MAGIC~\cite{MAGIC:2019fvw} while Fermi-LAT detected the source in the GeV-range~\cite{Fermi-LAT:2019yla,Fermi-LAT:2019pir}, and the overall \gr flux is at least an order of magnitude lower than expected in hadronic production models able to explain the observed \gr flux, unless invoking a large \gr absorption at the source \cite{blanco2023neutrinogammarayemissionngc}.
The lack of TeV \gr can be explained by \grs being emitted in the innermost vicinity of the supermassive black hole (SMBH; $R_{\rm em}\sim 10-100 \, R_S$, where $R_S$ is the Schwarzschild radius: $2GM_{\rm BH}/c^2$), 
which is an optically thick environment~\cite{Murase:2019vdl,Inoue_2020,Murase:2022dog,blanco2023neutrinogammarayemissionngc, Fiorillo:2023dts, Fiorillo_2024, Padovani:2024ibi}, together with an enhanced starburst activity farther away~\cite{Eichmann:2022lxh,Romeo:2016hms}. As a second explanation, it has been suggested that some beyond the Standard Model scenarios could also attenuate the \gr flux from \ngc, 
such as \gr oscillations into ALP, which subsequently decay into neutrinos~\cite{Pant:2023lnz}, or dark matter scattering with photons and neutrinos~\cite{Herrera:2025gpm,Cline:2023tkp}. 

The observation of high-energy neutrinos from \ngc led to dedicated searches of multi-messenger emission from other Seyfert galaxies~\cite{Neronov_2024,abbasi2024searchneutrinoemissionhard}. For instance, recent work by Ref.~\cite{Sommani:2024sbp} found two 100~TeV neutrinos that coincide with the Seyfert galaxy NGC~7469 at $3.3\,\sigma$ level, while no significant excess was found by Fermi-LAT. 
Other type-2 Seyferts are expected to be identified through ongoing IceCube observations, and upcoming data from IceCube-Gen2~\cite{Aartsen_2021}, KM3NeT~\cite{KM3Net:2016zxf}, P-ONE~\cite{Agostini_2020} and Baikal-GVD~\cite{gvdcollaboration2019neutrinotelescopelakebaikal}, combined with follow-up observations in \grs, X-rays, UV and optical. This motivates us to explore the multi-messenger spectrum in the context of ALPs. Moreover, given the excess of 79 neutrino events associated with \ngc, this source offers a unique opportunity for detailed investigation.

In this work, we focus on \ngc as a case study to describe its multi-messenger spectrum. We develop and discuss a jet model embedded in a thick corona to explain the \gr and neutrino emission. This model is consistent with observations and therefore enables us to search for dark matter induced modulations in the observed spectrum.
We consider, in particular, \gr conversion into ALPs in the magnetic fields along the jets, finding
$g_{a\gamma} \lesssim 7 \times 10^{-11}$GeV$^{-1}$ for $m_a \lesssim 10^{-9}$ eV at 95\% CL. These limits are slightly weaker than existing limits, however, when considering 10 \ngc-like sources, these limits improve by an order of magnitude, demonstrating the power of this recommended application in future analyses. 

This work is structured as following. In Sec.~\ref{sec:Part-I}, we describe the jet model, and we calculate the neutrino and \gr emission in the innermost vicinity of the central SMBH of \ngc. 
In Sec.~\ref{sec:Part-II}, we investigate the \gr conversion into ALPs in the magnetic fields along the jets, and in Sec.~\ref{sec:Part-II-Disc} we discuss the implications of our results.

\begin{figure*}[ht!]
    \centering
{{\includegraphics[width=0.9\textwidth]{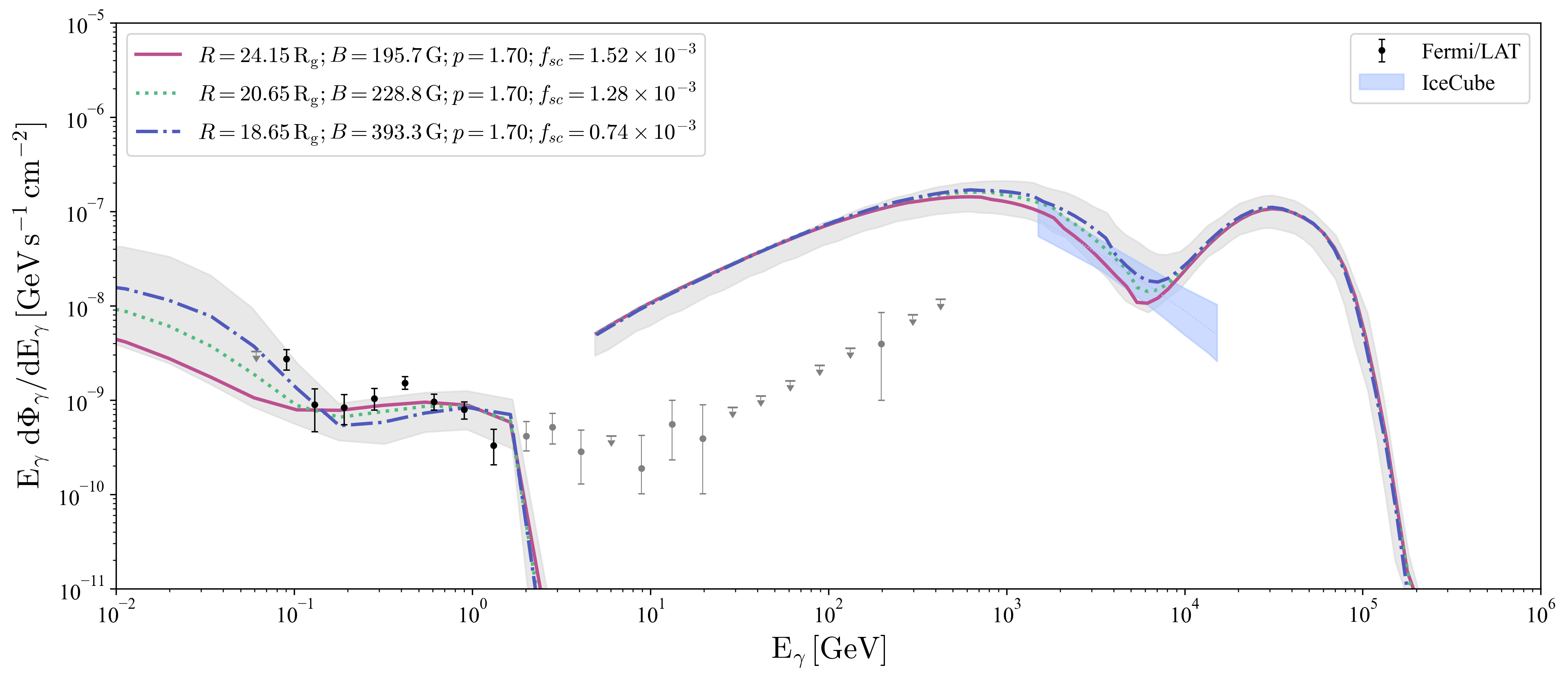} }}
  \caption{SED of \ngc focusing on the \gr regime and high-energy neutrinos, as measured by Fermi-LAT \cite{2012ApJ...755..164A, Fermi-LAT:2019yla} and IceCube \cite{IceCube:2022der}. The grey bands show the containment of the $10^{6}$ \hadjet simulations. We show 3 well-fitted individual models, where the pink line shows the best fit to the data. The resulting values of the free parameters $R,\, p,\,$\fsc and the calculated magnetic field strength from the magnetization $\sigma_f$ {as defined in eq.~\ref{eq: magnetisation}} are given in the legend. }\label{fig:SED_bestfit}
\end{figure*}

\section{Jet model}\label{sec:Part-I}
\subsection{Expected astrophysical \gr and Neutrino Fluxes from \ngc}\label{sec:Astro_models}

Radio interferometry of \ngc with the VLA and e-MERLIN has revealed a chain of compact components and a collimated jet extending from to kpc scales \cite{mutie2023radiojetsngc1068}. The jet features, together with the spatial coincidence between the jet and molecular outflows seen by ALMA provide evidence for a  large–scale jet capable of accelerating relativistic particles.

To calculate the \gr and neutrino emission from the accreting SMBH of \ngc, we adopt the jet model of \cite{lucchini2022BHJet}, referred to as \blj. This model accounts for the leptonic processes occurring along the jet, namely synchrotron emission of non-thermal electrons and inverse Compton scattering (ICS) with target photons, both originating in the jets and external photon fields. We further develop \blj to account for hadronic processes of non-thermal protons according to Ref.~\cite{kantzas2020cyg} and we refer to this model as \hadjet. We direct the interested reader to Refs.\cite{lucchini2022BHJet,kantzas2020cyg} for detailed descriptions, however, we briefly describe the model below. 

Two relativistic Poynting-flux dominated jets are launched by the accreting SMBH, perpendicular to the accretion plane, assuming that the two jets are identical. We assume that the jets are launched at a distance 2\,\rg (1\rg$\approx 1.5\,\rm km \,\times M_{BH}/M_{\odot}$) from the SMBH with a jet base of radius $r_0$. Whereas the bulk kinetic energy is carried by thermal protons, the thermal electrons at the jet base populate a Maxwell-J\"uttner (MJ) distribution in energies that peaks at 511\,keV. The thermal electrons emit synchrotron radiation that usually peak in the sub-keV regime. 

As the jets propagate, they accelerate up to some maximum Lorentz factor while the magnetic energy converts to kinetic. At distance \zdiss, the majority of the magnetic energy has dissipated to kinetic energy. At this distance, the jet magnetisation is determined by
\begin{equation}\label{eq: magnetisation}
    \sigma_f =\frac{B^2}{4\pi n m_{\rm p}c^2}<1,
\end{equation}
where 
$B$ is the magnetic field strength and $n$ is the particle number density \footnote{The geometry of the magnetic field in the coronal region of NGC 1068 is poorly constrained by radio observations \cite{mutie2025consistentradiosubmmpcscale, Michiyama:2023okx}. We consider the default jet-model geometry from \texttt{GammaAlps}, consisting of a toroidal component plus a turbulent component with the size of the corona $\sim 50R_g$.}. 
In this region, the majority of the power is carried by the particles. We assume that at \zdiss, electrons and protons start to accelerate to non-thermal energies driven by particle acceleration mechanism \citep{lucchini2022BHJet}. 
We remain agnostic to the specific mechanism and we capture this phenomenon by assuming that non-thermal particles follow a power in energies with index $p_e$, which is the same for both electrons and protons. We further parametrise the efficiency of the particle acceleration mechanism assuming the relative timescale \cite{jokipii1987rate}
\begin{equation}\label{eq: acceleration timescale}
    t_{\rm acc} (E) = \frac{4E}{3 f_{sc} {\rm ec} B},
\end{equation}
where $E$ is the particle energy, $e$ is the particle charge, $c$ is the speed of light and 
\fsc$<1$. Equating the acceleration timescale to the energy loss timescales, we calculate the maximum attainable energy of non-thermal electrons and protons along the jets. Finally, we assume that particles accelerate to non-thermal energies along the jets from \zdiss up to a maximum distance where the jets terminate in the intergalactic medium. This distance extends up to $10^8$\rg, hence does not affect the \gr and neutrino emission. We capture jet physics assuming that the jets consist of 100 consecutive segments on top of each other without any discontinuities from the jet base to the termination region, a sufficiently large number to allow for a multi-zone model without significantly reducing the speed of the numerical calculations. 

The non-thermal particles radiate in the entire multiwavelength spectrum. The leptonic particles emit synchrotron radiation that is further upscattered via ICS. Whereas these processes dominate the entire spectrum in many cases, for \ngc they are subdominant and they only dominate in the sub-GeV spectrum. 
The accelerated protons, on the other hand, interact with cold protons of the jet and the corona (see definition in the following paragraph), as well as the high-energy emission of the jet at the same jet segments, leading to proton-proton (pp) and photo-hadronic (\pg) processes, respectively. We calculate the resulting \gr and neutrino spectra following Ref.~\cite{kelner2006energy} for pp and Ref.~\cite{kelner2008energy} for \pg.

The SMBH with mass $M_{\rm BH}= 8 \times 10^{6} M_{\odot}$ \cite{Woo:2002un, Panessa:2006sg}
of \ngc is surrounded by a dense medium, known as the hot X-ray corona. 
Such a corona is responsible for the continuum observed in the X-ray spectra of type-1 Seyferts. This continuum is driven by the Compton scattering of thermal disk photons from the thermal corona electrons.
While thermal coronae are widely accepted to be the standard picture, non-thermal electron distributions have also been suggested for hot coronae that may irradiate the innermost accretion disk \citep{Asgari:2020hkq}. 
In Seyfert~1 AGN, several observational techniques indicate that the X-ray corona is extremely compact, typically within a few gravitational radii of the black hole. In particular, X-ray microlensing studies of lensed quasars constrain the half-light radius of the X-ray emitting region to 
$\sim$5–10R$_g$ \cite{Chartas:2008cg, Chartas:2015mta}.
For the case of \ngc, the corona is suggested to be extending to 30-100\rg to be optically thick in \grs, resulting in pair production, whereas the neutrinos produced by hadronic processes escape freely \cite{Murase:2022dog,Inoue_2020,Fiorillo_2024,karavola2025neutrino}. We choose to use a corona that extends at 50\rg allowing for a dense enough medium to absorb the \grs, and for an optical depth of $\tau=0.5$, the number density of the cold corona particles is $n=10^{11}\,\rm cm^{-3}$. The corona of \ngc acts as a target for the jet accelerated protons, namely the jet protons collide inelastically on the corona protons. To ensure that the particle acceleration inside the jet will occur in a region, which will be within the volume of the corona, we are only interested in values of \zdiss$<R_{\rm cor}=50$\rg. 
It should be noted that the number density of particles in the corona and its extent radius are uncertain, which translates into uncertainties on the gamma-ray opacity of the corona. The gamma-ray opacity is dominated by $\gamma \gamma \rightarrow e^{+}e^{-}$ and is proportional to $\tau \propto n R_{\rm cor}$. $n=10^{11}$cm$^{-3}$ is required for the region to be optically thick to Thomson scattering with cross section $\sigma_{\rm T}\simeq6.65 \times 10^{-25} \mathrm{~cm}^2$, which is the means by which the corona is defined. Fermi-LAT data constraints $R_{\rm cor} \lesssim 50R_g$, so the uncertainties on $\tau$ are within a factor of $\sim 10$ at most.

In principle, the starburst region of NGC 1068 may also contribute to the emitted gamma-ray flux. However, the gamma-ray luminosity observed by Fermi-LAT would require a starburst region with star-forming rate exceeding an order of magnitude the expectations of similar galaxies at such redshifts \cite{Lenain_2010}. It is thus plausible that the AGN/jet component provides the dominant gamma-ray emission contribution.

\subsection{Analysis}
To constrain the jet dynamics of \ngc and estimate the contribution of jet radiation in the high-energy regime, we fit
\hadjet to the Fermi-LAT data after selecting only models that reproduce IceCube measurements. In particular, we use the 15-year Fermi-LAT data from a $20^{\circ}\times 20^{\circ}$ region of interest surrounding \ngc between 50~MeV and 500~GeV \cite{2012ApJ...755..164A}, as has been analyzed by Ref.~\cite{blanco2023neutrinogammarayemissionngc}. 

In the Fermi-LAT analysis by Ref.~\cite{blanco2023neutrinogammarayemissionngc}, 14 energy bins are found that have both upper and lower limits on the energy spectrum. However, at the higher energy bins, an additional spectral contribution is expected from starburst activity farther away from the central engine, with supernova rates of $\sim 0.5 \,\mathrm{yr}^{-1}$ \cite{Eichmann:2022lxh} (see also Ref.~\cite{Ajello:2023hkh}), at least a factor of $\sim 5$ larger than expected from measurements of the Star Forming Rate \cite{1992ARA&A..30..575C} and supernova rates in similar galaxies \cite{VanBuren}. It is beyond the scope of this work to model the starburst contribution, and we therefore only select the first 8 bins in the analysis that are expected to have a jet origin. Indeed, at larger energies, the emitted \gr flux is efficiently attenuated by ambient photons in the corona, inducing a strong cut-off. 

We focus on two astrophysical free parameters that capture the jet morphology, namely the jet base radius $r_0$, and the magnetization $\sigma_f$ at the dissipation region \zdiss, which we set at 5\rg, well within the corona. To constrain the particle acceleration in the jets, we fit the power-law index of the non-thermal particles $p_e$ and the particle acceleration efficiency \fsc.  We then perform a $\chi^2$ statistical analysis to fit the Fermi-LAT data, considering only models that are found to produce a neutrino flux that fall within the observed uncertainty band from the IceCube measurements \cite{IceCube:2022der}. More precisely, we minimize
\begin{equation}\label{eq:chi2_SED}
\chi^2(r_0, \sigma_f, p_e, f_{\rm sc}) = \sum_{i}\frac{(\Phi_{\gamma,i}(r_0, \sigma_f, p_e, f_{\rm sc})-\Phi_{\gamma,i}^{\rm obs})^{2}}{(\delta \Phi_{\gamma,i}^{\rm obs})^{2}},
\end{equation}
where the index $i$ runs over the different bins in Fermi-LAT data from NGC 1068 in \cite{2012ApJ...755..164A,Fermi-LAT:2019yla}. 
We perform 10$^6$ simulations, varying the astrophysical parameters over the following motivated ranges: $r_0=[5, 30]$\rg, $\sigma_f=[0.01, 1]$, $p_e=[1.6, 2.4]$, and $f_{\rm sc}=[0.1, 20]\times 10^{-3}$.

\begin{table*}[ht!]
    \centering
    \setlength{\tabcolsep}{8pt} 
    \renewcommand{\arraystretch}{1.5} 
    \begin{tabular}{lclc}
        Parameter & Range of values &  Best-fit value & Description \\
        \hline
        $r_0$/\rg & [5, 30] & 24.15 & Jet base radius \\
        $\sigma_f$ & [0.01,1] &0.02 & Magnetization at dissipation region (Eq.~\ref{eq: magnetisation}) \\
        $p_e$ & [1.6, 2.4]&1.7 & Non-thermal power-law index $-\dfrac{\log(dn/dE)}{\log(dE)}$  \\
        \fsc & [0.1, 20] $\times 10^{-3}$ &$1.52\times 10^{-3}$ & Acceleration timescale factor \\
        \hline
        $\chi^2$ & $\geq23$ & 23 & Minimum $\chi^2$ fitted to Fermi-LAT data  \\
        \hline
        $n/\rm cm^{-3}$ & [4, 222]$\times 10^7$ & $10.13\times10^7$ & Particle number density at dissipation region  \\
        $B/\rm G$ & [157, 4709] & 196 & Magnetic field strength at dissipation region \\
    \end{tabular}
    \caption{Description of the astrophysical free parameters we consider in \hadjet and their ranges. Moreover, we show the best fit to \ngc Fermi-LAT data and the obtained $\chi^2$ values. The number density and the magnetic field are derived from the above parameters. }
    \label{tab:astro_parameters}
\end{table*}

\subsection{Results}\label{sec:Part-I-Results}
Figure~\ref{fig:SED_bestfit} shows the Spectral Energy Distribution (SED) of the best-fit model (solid purple line), both for \gr and neutrino emission. We further show in dashed and dotted lines other well-fitted solutions, and the gray shaded bands contain all our $10^{6}$ simulations of the expected \gr and high-energy neutrino emitted fluxes. 
First, we see that the emitting region has a cross-sectional radius of $\sim 20\,R_g$, well within the corona. The magnetic field is of the order of 200\,G and for a particle acceleration efficiency of $10^{-3}$, protons accelerate to TeV energies, allowing for the production of TeV neutrinos.
We moreover see that the Fermi-LAT spectrum can be well explained by pp \grs, whereas ICS contributes in the sub-GeV regime. The emitted \gr spectrum is heavily attenuated by the corona, whereas the produced neutrinos propagate freely. 
In particular, the emitted neutrino spectrum has a double-bump shape due to the contribution of both pp (first peak) and \pg neutrinos (second peak). 
We show the total neutrino and anti-neutrino spectrum per neutrino flavor and we compare it to the observed neutrino spectrum by IceCube (shaded blue region).
The emitted neutrino signal falls well within the observed limit, with some minor excess in the $\sim$10\,TeV regime. Further observations could possibly prove this feature or constrain it. To calculate the event rate, we follow Ref.~\cite{kantzas2023possible}, accounting for the effective area of IceCube for zenith angles between -5 and +30 degrees. We find that 92 muon neutrino events are expected, in reasonable agreement with the IceCube reported number of $79\pm20$ events \cite{IceCube:2022der}.

\section{ALP-photon oscillation}\label{sec:Part-II}
The coupling of ALPs to photons is described by the Lagrangian,
\begin{equation}
\mathcal{L}_{a \gamma}=-\frac{1}{4} g_{a \gamma} F_{\mu \nu} \tilde{F}^{\mu \nu} a,
\end{equation}
where $F_{\mu \nu}$ is the electromagnetic field tensor, $\tilde{F}_{\mu \nu}$ its dual, and $a$ is the ALP field strength. The effective Lagrangian for the ALP-photon system is 

\begin{equation}
\mathcal{L}=\mathcal{L}_{a \gamma}+\mathcal{L}_{\mathrm{EH}}+\mathcal{L}_a,
\end{equation}
where $\mathcal{L}_{\mathrm{EH}}$ is the Euler-Heisenberg Lagrangian accounting for one-loop corrections to the photon propagator, and the ALP Lagrangian terms are

\begin{equation}
\mathcal{L}_a=\frac{1}{2} \partial_\mu a \partial^\mu a-\frac{1}{2} m_a^2 a^2 .
\end{equation}

ALPs couple to photons in the presence of a magnetic field component transversal to the direction of propagation. The equations of motion for a photon beam with energy E read \cite{Raffelt_stodoslky, DeAngelis:2011id}

\begin{equation}
\left(i \frac{\mathrm{d}}{\mathrm{d} z}+E+\mathcal{M}\right) \Psi\left(z\right)=0,
\end{equation}
with $\Psi\left(z\right)=\left(A_x\left(z\right), A_y\left(z\right), a\left(z\right)\right)^T$, A's being the polarization states along directions x and y and the mixing matrix $\mathcal{M}$ reads
\begin{equation}
\mathcal{M}=\left(\begin{array}{ccc}
\Delta_{\perp} & 0 & 0 \\
0 & \Delta_{\|} & \Delta_{a \gamma} \\
0 & \Delta_{a \gamma} & \Delta_a
\end{array}\right).
\end{equation}

QED vacuum polarization effects and the propagation of photons in the plasma medium give rise to the terms $\Delta_{\perp}=\Delta_{\mathrm{pl}}+2 \Delta_{\mathrm{QED}}$ and $\Delta_{\|}=\Delta_{\mathrm{pl}}+7 / 2 \Delta_{\mathrm{QED}}$, 
with $\Delta_{\mathrm{pl}}=-\omega_{\mathrm{pl}} /(2 E)$ and the plasma frequency $\omega_{\mathrm{pl}} \sim 0.037 \sqrt{n})$, where $n$ is the number density of electrons in the plasma. 
The $\mathrm{QED}$ vacuum polarization term reads $\Delta_{\mathrm{QED}}=\alpha E /(45 \pi)\left(B /\left(B_{\mathrm{cr}}\right)\right)^2$, with the fine-structure constant $\alpha$, and the critical magnetic field $B_{\mathrm{cr}}=m_e^2 /|e| \sim 4.4 \times 10^{13} \mathrm{G}$. 
The kinetic term for the ALP is $\Delta_a=-m_a^2 /(2 E)$ and photon-ALP mixing is given by the off-diagonal elements $\Delta_{a \gamma}=g_{a \gamma} B / 2$. 
The above system of equations can be solved with the software \texttt{GammaALPs} for a combination of astrophysical environments, which allows obtaining the photon survival probability as a function of energy~\cite{Meyer:2021pbp}. From this quantity, the axion field opacity to \grs can be derived. 
As described in the previous section, we consider a jet-based mechanism for the acceleration of cosmic rays and production of high-energy neutrinos and \grs. Thus, we consider the \texttt{GammaALPs} ``Jet'' model, where the ALP-photon mixing occurs in the toroidal magnetic field of the AGN jet. In this case, the magnetic field is coherent, and modeled as a power-law in distance from the central black hole as
\begin{equation}
B(r)=B_0\left(\frac{r}{R_{\mathrm{em}}}\right)^{-1},
\end{equation}
where $R_{\rm em}$ is the emitting region of \grs of the AGN. Similarly, the electron number density is modelled as a power-law function
\begin{equation}
n(r)=n_0\left(\frac{r}{R_{\mathrm{em}}}\right)^{-2} .
\end{equation}

For the attenuation of \grs in the Extragalactic Background Light, we use the model from Ref.~\cite{Dominguez_2010}. We do not include the intergalactic magnetic field, as its field strength is too weak to have any impact on ALP-photon oscillations for the parameters considered in this work (see also Appendix~\ref{app:GMF}). 
Finally, we also consider the possibility of ALP-photon re-conversions in the coherent component of the magnetic field of the Milky Way. In that case, the Galactic magnetic field (GMF) model is taken from Refs.~\cite{Pshirkov_2011, Jansson_2012} with constant number density of electrons~\cite{Horns:2012kw}and we discuss uncertainties in the GMF in Appendix~\ref{app:GMF}.

Under this prescription, we can compute the attenuated \gr flux due to ALP-photon oscillations as 
\begin{equation}
    \frac{d\Phi_{\gamma}^{\rm att}}{dE_{\gamma}}=\frac{d\Phi_{\gamma}}{dE_{\gamma}} ~P_{a\gamma}(E_{\gamma},m_a,g_{a\gamma},\boldsymbol{\theta}_{\rm astro}),
\end{equation}
where $\Phi_{\gamma}^{\rm att}$ is the attenuated \gr flux and $\Phi_{\gamma}$ the expected \gr flux from the accreting SMBH of \ngc. $P_{a\gamma}$ is the ALP-photon oscillation that depends on the \gr energy $E_{\gamma}$, axion mass $m_a$, axion-photon coupling $g_{a\gamma}$, and the range of astrophysical parameters $\boldsymbol{\theta}_{\rm astro}$ as discussed in the next section.

\subsection{ALP analysis}\label{sec:Part-II-Analysis}
We derive limits on the ALP parameter space, and identify ALP regions that better fit the Fermi-LAT and IceCube data. 
We perform a profile likelihood analysis by marginalizing over the astrophysical nuisance parameters $\boldsymbol{\theta}_{\rm astro}=(r_0, \sigma_f, p_e, f_{\rm sc})$ that predict different SED, as illustrated in Fig.~\ref{fig:SED_bestfit} as a gray band. For each value $(m_a, g_{a\gamma\gamma})$, we compute $\chi^2(m_a, g_{a\gamma\gamma};\boldsymbol{\theta}_{\rm astro})$ by minimizing over the nuisance parameters, where $\chi^2(m_a, g_{a\gamma\gamma};\boldsymbol{\theta}_{\rm astro})$ is obtained through a least-squares analysis using the Fermi-LAT data of NGC 1068. 
We obtain 95\% CL limits on $g_{a\gamma\gamma}$ as a function of $m_a$ through a Likelihood-Ratio analysis, which follows Ref.~\cite{Wilks:1938dza}
\begin{equation}\label{eq:chi2_ALP}
\chi^2(m_a, g_{a\gamma\gamma};\boldsymbol{\theta}_{\rm astro})-\chi^2_{\rm min}(\boldsymbol{\theta}_{\rm astro}) = \Delta \chi^2  \leq 2.71.
\end{equation}
The considered ranges for the nuisance parameters are shown in Table \ref{tab:astro_parameters}. 

\begin{figure}[t!]
    \centering \label{fig2}
{{\includegraphics[width=0.5\textwidth]{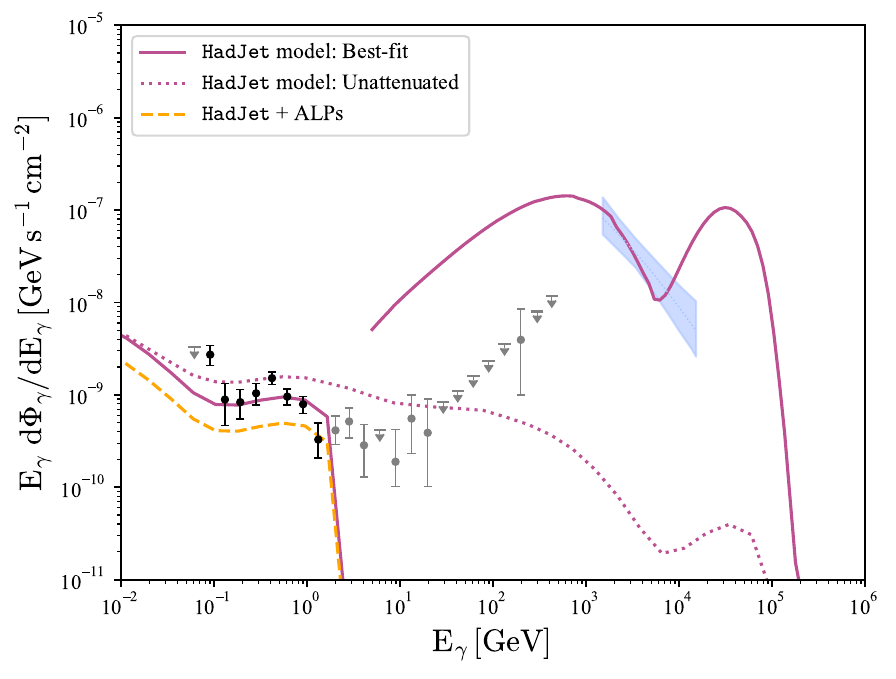} }}
  \caption{SED of \ngc, including \gr measurements from Fermi-LAT \cite{2012ApJ...755..164A, Fermi-LAT:2019yla}, high-energy neutrino measurements from IceCube~\cite{IceCube:2022der}, and the best-fit model (solid purple line). 
  The opaque \gr flux due to ALP-photon oscillations in the jet of \ngc is shown as orange dashed line with ALP parameters $m_a=10^{-7}$~eV and $g_{a\gamma}=10^{-10}$~GeV$^{-1}$. 
  }\label{fig: NGC1068 multimessenger spectrum with best fit solution}
\end{figure}
\begin{figure}[t!]
    \centering
{{\includegraphics[width=0.535\textwidth]{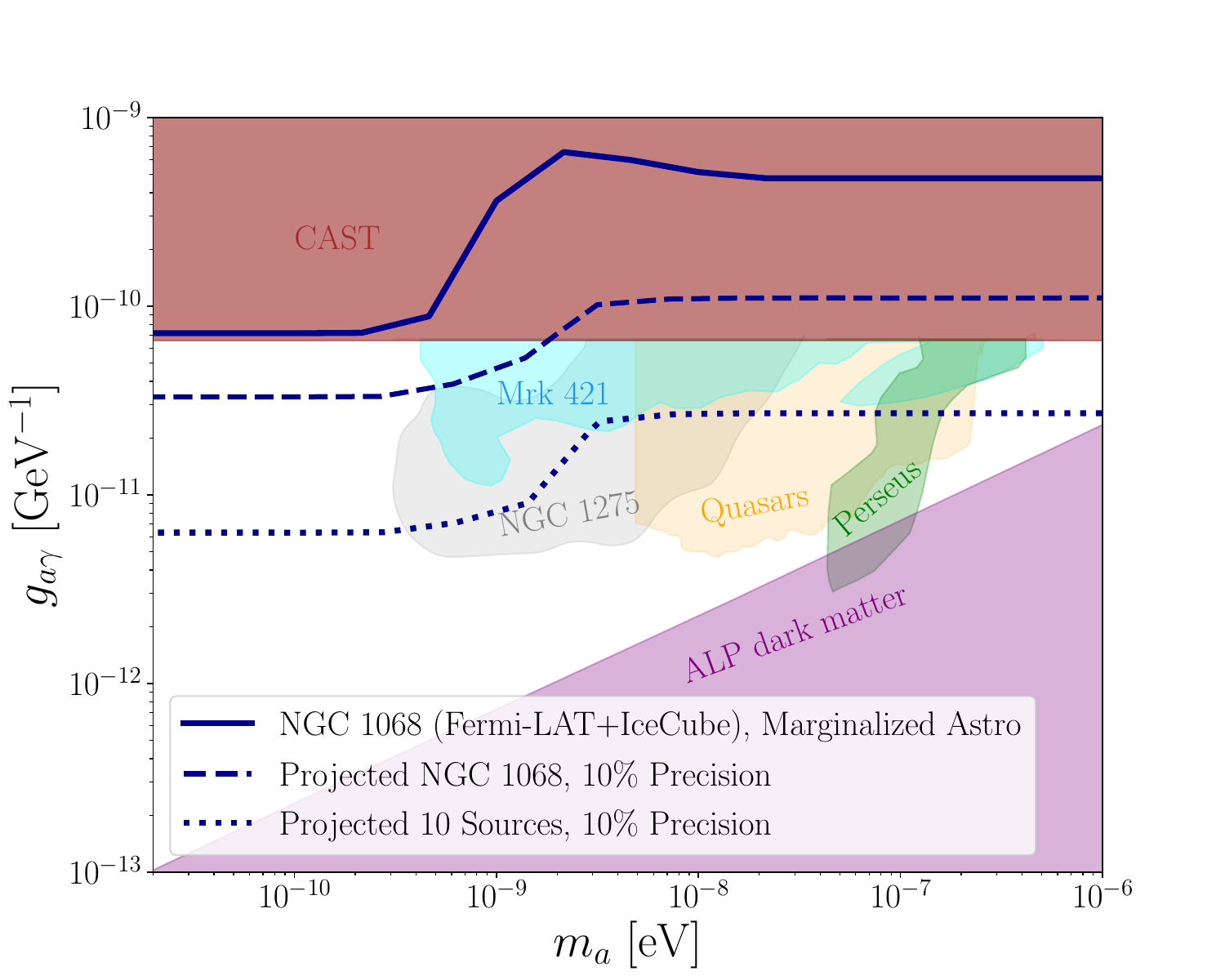} }}
  \caption{Upper limits on the ALP-photon coupling versus ALP mass by marginalizing over astrophysical parameters described in Table~\ref{tab:astro_parameters} (solid blue line). We further show projected limits with improved Fermi-LAT precision of 10$\%$ (dashed blue line), and from the observation of 10 \ngc-like sources with 10$\%$ precision (dotted blue line). We compare with previous results~\cite{Li:2020pcn, Fermi-LAT:2016nkz, Quasars, Abe_2024}, and show the region where ALPs have the correct relic abundance via the misalignment mechanism (shaded purple)~\cite{Essig:2013lka}.}

  \label{fig:upper_limit}
\end{figure}

\subsection{Results}\label{sec:Part-II-Results}
Figure~\ref{fig: NGC1068 multimessenger spectrum with best fit solution} shows the best-fit expected \gr and neutrino emission in solid purple lines, together with the unattenuated emission (dotted purple line) for illustration purposes. We further show in a dashed orange line the additionally attenuated \gr flux induced by ALP-photon oscillation effects in the magnetic field of the jets of \ngc, for benchmark values of the ALP mass $m_{a}=10^{-7}$~eV and axion-photon coupling $g_{a\gamma}=10^{-10}$~GeV$^{-1}$. 
We see that the ALP-photon oscillation can sizably deplete the \gr flux below observational levels. This feature can be used to set limits on the ALP-photon coupling and mass. 

We find a best-fit point in the ALP parameter space at $(m_a, g_{a\gamma \gamma})=(2.2\times 10^{-9}~\text{GeV}^{-1}, ~4.6\times 10^{-10}~\text{eV})$, corresponding to a significance of only $1.02 \sigma$, indicating no meaningful preference.

In Fig.~\ref{fig:upper_limit}, we show the ALP-photon coupling versus the ALP mass.
The solid blue line corresponds to the 95\% CL limits obtained when marginalizing over the astrophysical parameters. 
Moreover, we compare our limits to previous bounds arising from ALP-photon oscillations in $\gamma$-ray astrophysical sources. In particular, we show bounds from the combination of ARGO-YBJ and Fermi-LAT observations of the blazar Mrk 421 (dashed cyan region) \cite{Li:2020pcn}, Fermi-LAT observations of NGC 1275 (dashed gray region) \cite{Fermi-LAT:2016nkz}, flat-spectrum radio quasars observed with Fermi-LAT (dashed orange region) \cite{Quasars}, and observations of the Perseus Galaxy with MAGIC (dashed green region) \cite{MAGIC:2019fvw}. We also show the bound obtained by the CAST experiment at CERN, based on solar ALP-photon conversions in the presence of a strong magnetic field in the laboratory \cite{CAST:2017uph}. 

It should be noted that additional limits exist on the ALP-photon coupling from ALP production and reconversion into X-rays
from ALP-induced polarization effects, and from reionization probes, \textit{e.g} \cite{AxionLimits, Li:2020pcn,Pant:2023lnz,2024arXiv240519393M,Dessert:2022yqq,Noordhuis:2022ljw, Escudero:2023vgv, Eckner:2022rwf,BetancourtKamenetskaia:2024gcv,Benabou:2025jcv,Ning:2024eky,Dessert:2020lil}. We further show the region of parameter space where ALPs can constitute the dark matter of the Universe (shaded purple region) \cite{Essig:2013lka}. Our current limits do not yet reach this ALP dark-matter target region in the mass range where our multimessenger bounds are most competitive, but the projected sensitivities shown in Fig.~\ref{fig:upper_limit} illustrate how improved measurements and a population of NGC~1068-like sources could move closer to that parameter space.

Figure \ref{fig:upper_limit} also illustrates projected limits from NGC 1068 with improved precision of 10$\%$ in the measurements (dashed blue line), and projected limits with the observation of 10 \ngc-like sources and 10$\%$ precision in each of the measurements (dotted blue line). 
This would constitute an improvement w.r.t to current measurements from Fermi-LAT, which present uncertainties in the range $16.7\%-50\%$, depending on the energy bin considered. We generate mock astrophysical predictions and data by performing a 1$\sigma$ Gaussian perturbation of the best-fit astrophysical flux from \ngc discussed in Sec.~\ref{sec:Part-I}, together with the central values of the measurements of Fermi-LAT data. We note that using the best-fit astrophysical model instead of marginalizing over astrophysical parameters improves the limits by $<20\%$. The mock astrophysical fluxes and corresponding observed data are then generated as
\begin{equation}
\Phi^{\rm mock (\rm obs)}_{\gamma,i}\left(E_i\right)=\Phi_{\gamma,i}^{(\rm obs)}\left(E_i\right) \times\left(1+\delta_i\right), \, \delta_i \sim \mathcal{N}(0,1),
\end{equation}
where $\mathcal{N}(0,1)$ denotes a Gaussian distribution centered at $\mu=0$ with standard deviation $\sigma=0.1$. Under this prescription, and fixing the uncertainty in mock data to 10$\%$ in every energy bin, we find that our limits could significantly improve previous limits from \gr sources for some values of the ALP mass. We note that measurement uncertainties of 10\% have been obtained for other Fermi-LAT sources in the 20-500 GeV energy-range, and future experiments like AMEGO \cite{Caputo:2022xpx}, e-Astrogram \cite{e-ASTROGAM:2017pxr} or CTAO \cite{CTAConsortium:2017dvg} are expected to improve over Fermi-LAT precision.

\section{Discussion}\label{sec:Part-II-Disc}
The multi-messenger era brings us the opportunity to understand the underlying processes leading to the observed \gr and high-energy neutrino fluxes from distant astrophysical sources. In this context, beyond the Standard Model scenarios may induce modifications of the emitted astrophysical \gr or neutrino fluxes, leaving the former or the latter unaltered. ALP-photon oscillations are widely discussed to modify the \gr flux from extragalactic sources, like AGNs. 
We propose to use multi-messenger sources that are observed in high-energy neutrinos, as they offer a unique opportunity to disentangle intrinsic emission mechanism from ALP-induced effects. This enables to derive robust limits on the ALP-photon coupling. 

For this purpose, we develop a model that allows for lepto-hadronic processes between jet accelerated protons and cold target particles of the corona. Using this model, we calculate the expected \gr and high-energy neutrino fluxes from \ngc, and compare to IceCube and Fermi-LAT data. 
The observation of high-energy neutrinos by IceCube significantly constrains the astrophysical parameter space allowing for more robust conclusions on the origin of the non-thermal emission.
We find that ICS of jet accelerated electrons can reproduce the sub-GeV spectrum, pp inelastic collisions between accelerated jet protons and corona thermal protons can contribute or even dominate the GeV spectrum, whereas neutrinos from both pp and \pg interactions can explain the $\sim$80 events detected by IceCube. The entire high-energy \gr emission is suppressed due to the optically thick corona.

Using these simulations, we derive limits on the ALP-photon coupling, at the level of $g_{a\gamma} \lesssim 7 \times 10^{-11}$GeV$^{-1}$ for $m_a \lesssim 10^{-9}$ eV. These limits are weaker than previous analyses of astrophysical \gr sources. 
However, to illustrate the potential of our method in future analyses, we expect that with a population of 10 \ngc-like sources and considering an improved precision in future \gr measurements of 10$\%$, to find comparable and complementary limits when compared to current results from \gr sources.

Our analysis can be extended in several ways. One may consider different \gr sources, such as TXS or PKS blazars, which have also been observed in \grs and high-energy neutrinos \cite{IceCube:2018dnn}. Furthermore, other Seyfert galaxies have been observed by IceCube with moderate significance \cite{abbasi2024searchneutrinoemissionhard}, and while Fermi-LAT currently does not find a significant \gr flux, future observations could improve the limits. 
Beyond the Standard Model scenarios could be considered as attenuating media in these sources. 
We expect future multi-messenger and multi-wavelength observations to help determine the dark opacity of the Universe to \grs.

\section{Acknowledgments}
We thank Francesca Calore and Marco Chianese for valuable comments on the draft. 
AD is supported by the Kavli Institute for Cosmological physics at the University
of Chicago through an endowment from the Kavli Foundation and its founder Fred Kavli. 
The work of GH is supported by the U.S. Department of Energy under award number DE-SC0020262.
DK acknowledges funding from the French Programme d’investissements d’avenir through the Enigmass Labex, from the ‘Agence Nationale de la Recherche’, grant number ANR-19-CE310005-01 (PI: F. Calore), and from Tamkeen under the NYU Abu Dhabi Research Institute grant CASS. 

\appendix
\section{Galactic magnetic field}\label{app:GMF}
\begin{figure}[h]
\centering
\includegraphics[width=0.49\textwidth]{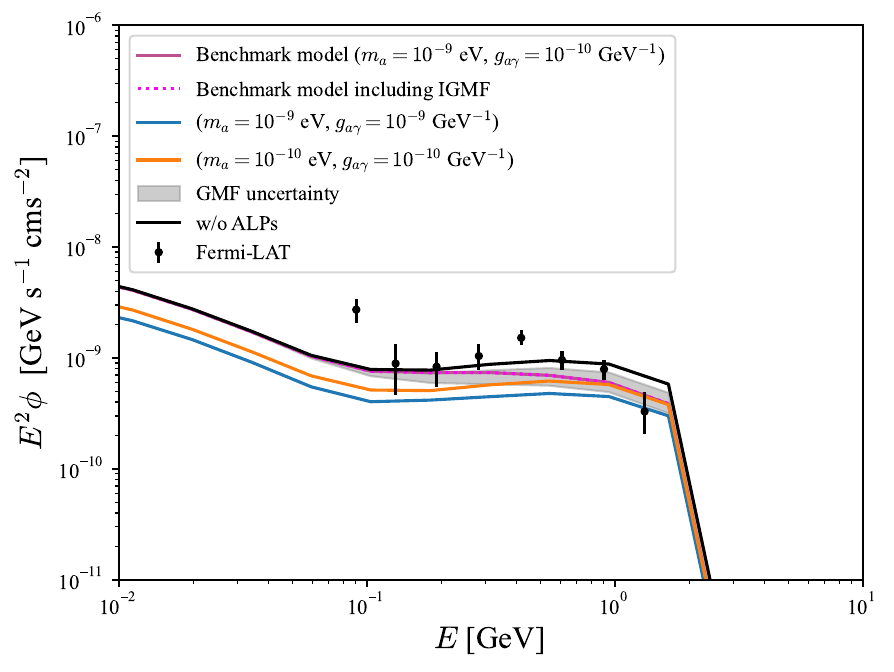}
\caption{SED of NGC 1068, including \gr measurements from Fermi-LAT (black error bars). The best-fit model without ALPs is shown by the black line. The pink line shows the spectrum including attenuation from ALP-photon oscillations for $m_a=10^{-9}$~eV and $g_{a \gamma}=10^{-10}$~GeV$^{-1}$, assuming the benchmark GMF model~\cite{Jansson_2012}. The grey envelope illustrates the uncertainties from eight different models~\cite{Unger:2023lob}.}
\label{fig:GMF}
\end{figure}

The Galactic magnetic field (GMF) is subject to uncertainties in the field morphology and in the thermal electron density. As a benchmark model, we adopt the GMF model of Ref.~\cite{Jansson_2012}. To assess the impact of these uncertainties, we repeat the analysis using the eight updated GMF models presented in Ref.~\cite{Unger:2023lob}, that are implemented in \texttt{GammaALPs}.
The result is shown in Figure~\ref{fig:GMF} as a grey envelope, while the benchmark model including ALP is indicated as the pink line ($m_a=10^{-9}$~eV, $g_{a\gamma}=10^{-10}$~GeV$^{-1}$) and the best-fit astrophysical model as black line. We find that the corresponding 95\% CL limits on $g_{a \gamma}$ are only affected for masses below $m_a \lesssim 10^{-8}$~eV and weaken by at most 56\% and strengthened by 37\% relative to the benchmark model. The orange and blue line show different ALP models, illustrating the effect of changing the mass and coupling strength. 

Moreover, the dotted line shows the case including the IGMF (with $\sim 10^{-12}$ G and $\lambda\sim$ Mpc) where it is clear that it has no effect on the ALP-photon oscillation in the parameters considered here.

\bibliography{References}

\end{document}